\documentstyle[twocolumn,cite,epsf]{elsart}
\def\ket#1{\left | {#1} \right >}
\def\bra#1{\left < {#1} \right |}
\def\avg#1{\left < {#1} \right >}

\def\pr{ {\em Phys. Rev.}}

\def\prb{{\em Phys. Rev. {\bf B} }}
\def\prl{{\em Phys. Rev. Lett. }}

\def\mplb{{\em Mod. Phys. Lett. {\bf B}}}
\def\beq{\begin{equation}}
\def\eeq{\end{equation}}

\begin{document}
\small
\begin{center}
\begin{frontmatter}

\title{Successive crossover from ordinary Born scattering to multiple
scattering to localization - A delay time analysis in electronically
random systems}
\author{Sandeep K. Joshi\cite{jos} and A. M. Jayannavar\cite{amje}}
\address{$^*$Institute of Physics, Sachivalaya Marg, Bhubaneswar 751 005,
India}

\maketitle

\begin{abstract}

We have studied the reflection delay time distribution from a
one-dimensional electronically random system of finite length. We show
that the average delay time is a non-monotonic function of the strength of
the disorder and exhibits three qualitatively different regimes. In the weak
disorder limit the scattering is dominated by the ordinary Born
scattering. For the intermediate strengths of disorder a well defined regime
of multiple scattering emerges followed by the localization domain.

\end{abstract}
\end{frontmatter}
\end{center}

\section{INTRODUCTION}
\label{intro}

Wigner delay time, the energy derivative of scattering phase shift, is
considered to be related to the time spent by the particle in the region
of interaction\cite{wigner,smith}.  Since its inception in 1955 the Wigner
delay time has been a quantity of profound interest from fundamental as
well as technological point of view
\cite{thomas,gopar,nku,butt+}. The delay time statistics is
intimately connected with the dynamic admittance of
microstructures\cite{thomas}. Although the phase delay time alone is not
sufficient to calculate capacitance of extended mesoscopic
systems, a direct relation between the capacitance and the
phase time does exist for systems which can be treated essentially as zero
dimensional \cite{butt+}(eg. a cavity or a quantum dot). The wave packet
incident on the surface of a sample is backscattered after some delay. In
random systems this results in a low temperature $1/f$ type noise for the
fluctuating surface currents due to non-cancellation of the instantaneous
currents at the surface\cite{amj}. For a three dimensional system of
arbitrary shape with an arbitrary number of incoming channels the density
of states is proportional to the sum of dwell times for all incoming
channels\cite{ianna}. In case of a one dimensional disordered sample
connected to an infinite lead the dwell time is the same as the Wigner
delay time. This delay time is directly related to the density of states.

In the single channel case the distribution of the delay time for a
disordered semi-infinite sample was obtained earlier by using the
invariant imbedding approach\cite{amj}. The stationary distribution
$P_s(\tau)$ for the dimensionless delay time $\tau$ is given by

\beq
P_s(\tau)~=~\frac{\lambda e^{\lambda
tan^{-1}\tau}}{(e^{\lambda\pi/2}-1)(1+\tau^2)},
\label{pst}
\eeq

\noindent where $\lambda$ is proportional to the disorder induced
localization length and the most probable value of $\tau$ occurs at
$\tau_{mp}=\lambda/2$. The long time tail of the above distribution scales
as $1/\tau^2$. The average value of $\tau$ is logarithmically divergent
indicating the possibility of the particle traversing the infinite sample
before being totally reflected, due to the resonances
\cite{texier,epstein}.  If the disordered region is semi-infinite or if it
is terminated at one end by an infinite wall, the reflection coefficient
will be unity, and the complex reflection amplitude will have the form
$R~=~e^{i\theta(E)}$.

A recent study based on analytical work found the delay time distribution
in the one-channel case to be universal (especially the long time tail is
independent of the nature of disorder)\cite{comtet} within random phase
approximation (RPA). Soon after the universality of tail distribution
beyond RPA was established numerically\cite{joshi-prb}. The universality
of stationary distribution of delay time was further supported by another
analytical work where the random potential is of a different
kind\cite{joshi-ssc}. So far in this field the localization has been
studied extensively.  Recently Texier et. al. \cite{texier} have studied
the statistics of delay time distribution for ballistic case for which
sample size is much smaller than the localization length. In the present
paper we analyse the statistical properties of delay time and study the
emergence of three qualitatively different regimes of transport or
scattering in a finite one-dimensional random system. These different
domains are dominated by ordinary Born scattering, multiple scattering and
localization respectively. The crossover from one regime to
another as the disorder strength is varied is studied.

\section{ NUMERICAL TECHNIQUE}
\label{numerics}

The following Hamiltonian, written in a tight-binding, one-band model is
used to describe the motion of a quasiparticle on a random lattice:

\beq
H~=~\sum~\epsilon_n\ket{n}\bra{n}~+~V(\ket{n}\bra{n+1}~+~\ket{n}\bra{n-1}),
\label{ham}
\eeq

\noindent where $\ket{n}$, $\epsilon_n$ and $V$ denote the non-degenerate
Wannier orbital at site $n$, the site energy at the site $n$ and the
hopping matrix element connecting nearest neighbors separated by a unit
lattice spacing respectively. We consider site diagonal disorder where the
site energies $\epsilon_n$ are assumed uncorrelated random variables
distributed uniformly over $-W/2$ to $W/2$ ($P(\epsilon_n)=1/W$). The $L$
site ($n~=~1~to~L$) disordered 1D sample is connected at one end to a
perfect infinite lattice having all site energies zero. The well-known
transfer-matrix method\cite{trm,econ} is used to calculate the reflection
amplitude $r(E)~=~|r|e^{-i\theta(E)}$ and its phase $\theta(E)$ at two
values of incident energy $E=E_0\pm\delta E$. The delay time is then
calculated using the definition $\tau = \hbar d\theta/dE$.

Throughout our following discussion we consider the delay time $\tau$ in a
dimensionless form by multiplying them with $V$ and we set $\hbar=1$.
For all the results discussed below we have used $E_0=0$ and $\delta
E=10^{-8}$. We have confirmed the stability of our results with respect to
choice of $\delta E$ within the range $10^{-6}~<~\delta E~<~10^{-9}$. For the
calculation of various distributions and averages we have used atleast
$10^6$ realizations of the disordered sample.

\section{Results and discussion}
\label{results}

\begin{figure}[t]
\protect\centerline{\epsfxsize=3in \epsfbox{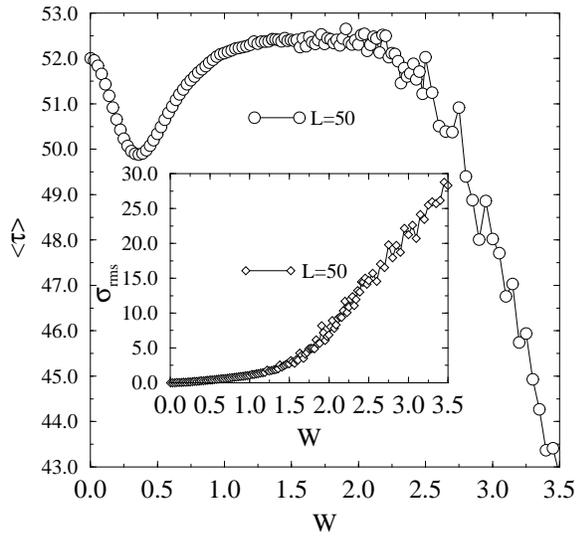}}
\caption{Plot of $\avg{\tau}$ versus disorder strength $W$ for different
lengths of the disordered segment. The disorder averaging is done over
$10^7$ samples.}
\label{avtau-W}  
\end{figure}

In Fig. \ref{avtau-W} we show the plot of average value of delay time
$\avg{\tau}$ versus the disorder strength for a sample of length $L=50$.
For very small values of $W$ such that localization length $\xi \gg L$
($\xi=96/W^2$ \cite{econ}), we see that the $\avg{\tau}$ decreases with
increasing $W$. This is the ballistic regime. Since the disorder is very
weak i.e., $\xi \gg L$ Born scattering dominates. Due to this the waves
undergo mostly single scattering events. The $\avg{\tau}$ continues to
fall upto a certain strength of disorder depending upon the length of the
sample. At this strength of disorder $\xi\sim 20L$. We have verified this
for samples of different lengths. Increasing $W$ beyond this value we
observe that $\avg{\tau}$ starts increasing with disorder, i.e., the
disorder becomes strong enough such that multiple scattering events start
dominating. This leads to particles (or waves) spending more time in the
sample before getting reflected. Consequently, the reflection delay time
increases with increasing disorder. This trend continues until the
disorder becomes strong enough such that $\xi \sim L$. On increasing $W$
beyond this, a crossover to localization occurs. In this regime the most
probable distance traversed by the particles is twice the localization
length. The localization length decreases with disorder strength as
$96/W^2$. Therefore, the most probable value of the reflection delay time,
the time required to travel to-and-fro distance of $\xi$, also decreases
and so does $\avg{\tau}$. However, from the inset shown in the figure we
see that the root mean squared relative variance
$\sigma_{rms}=\sqrt{\avg{\tau^2}-\avg{\tau}^2}/\avg{\tau}$ in this regime
is much larger than unity. This indicates that the fluctuations in $\tau$
in this regime are very large and hence $\tau$ is a non-self-averaging
quantity. Thus, from the behaviour of average reflection delay time
$\avg{\tau}$ we are able to identify three qualitatively different regimes
of wave propagation in one-dimensional disordered medium.

\begin{figure}[t] 
\protect\centerline{\epsfxsize=3in \epsfbox{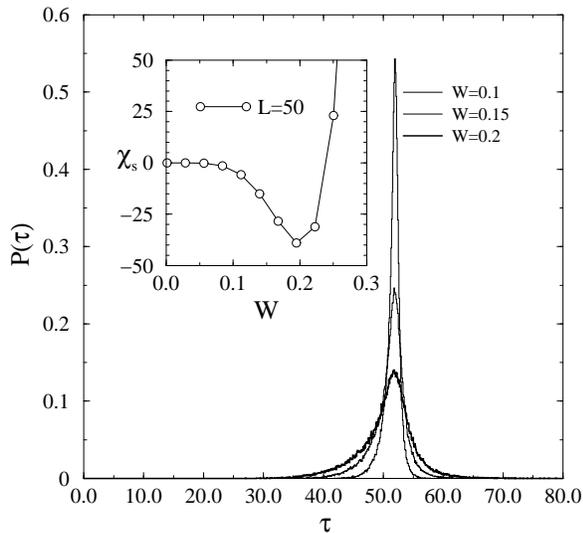}} 
\caption{Distribution of $\tau$
from a sample of length 50 for three different values of disorder strength
$W$ for the ballistic regime $\xi \ll L$. The inset shows the behaviour of
third moment of $\tau$ with $W$.}
\label{tau-dist1}
\end{figure}

In Fig. \ref{tau-dist1} is shown the distribution $P(\tau)$ of reflection
delay time $\tau$ from a sample of fixed length $L=50$ for three values of
disorder strength $W$ in the ballistic regime. For $W=0$ the distribution
is a delta peak at a value of $\tau$ equal to the round trip delay time
$\tau_0$. For the weak disorder case the most probable value of $\tau$ is
the round trip delay time $\tau_0$. From the figure we observe that the
distributions are asymmetric about the most probable value $\tau_0$. The
distribution has higher weight for values $\tau ~<~ \tau_0$ and lower
weight for values of $\tau ~>~ \tau_0$. The asymmetry is confirmed by
looking at the third moment of $\tau$ which is $\chi_s~=~\avg{(\tau -
\avg{\tau})^3}$ shown in the inset of Fig. \ref{tau-dist1}. The variation
of $\chi_s$ with $W$ sheds some light on the behaviour of $\avg{\tau}$
with $W$ in the ballistic regime. In this regime, as mentioned earlier,
disorder is weak and much of the behaviour is dominated by single
scattering events which occur anywhere in the sample before reaching
endpoint. As the disorder strength is increased the probability of
scattering at individual sites also increases. Thus, the probability of
particle encountering a scattering near the surface itself increases with
increasing disorder strength. This enhances the weight for small $\tau$ in
$P(\tau)$. Although, the multiple scattering events contribute to the
distribution for $\tau ~>~ \tau_0$, the change in their contribution with
increasing disorder is not much compared to that of single scattering
events.

\begin{figure}[t] 
\protect\centerline{\epsfxsize=3in \epsfbox{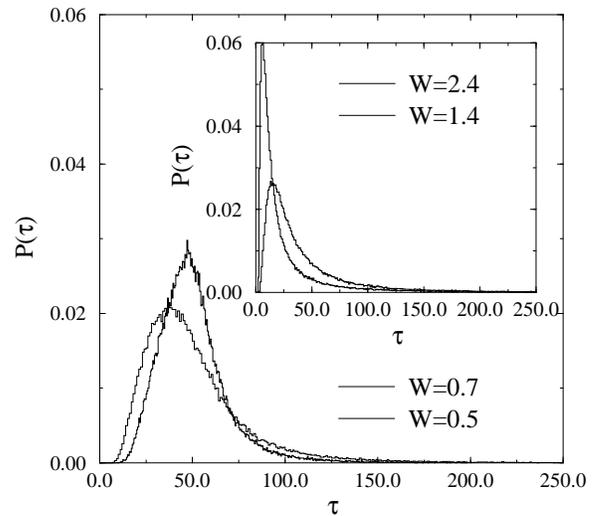}} 
\caption{Distribution of $\tau$
from a sample of length $L=50$ for different values of disorder strength
$W$. For $L=50$, $W=0.5$ and $W=0.7$ lie in the multiple scattering regime.
Inset shows the distribution of $\tau$ for $W=1.4$ and $W=2.4$ belonging 
to the localization regime. } 
\label{tau-dist2}
\end{figure}

The distributions of delay time in the multiple scattering regime and the
localized regime are shown in Fig. \ref{tau-dist2}. For the distribution
in the multiple scattering regime we notice that the most probable delay
time $\tau_{mp}$ is much lower than the round trip delay time $\tau_0$.
The distribution is broader and has a tail. The appearance of a broad
distribution and a long time tail indicates that in this regime the
multiple scattering events are indeed dominating. In fact, it is due to
the appearance of a tail that the $\avg{\tau}$ increases with increasing
disorder strength $W$ in spite of the reduction in $\tau_{mp}$. For $W ~>~
\sqrt{2}~(\xi ~<~ L)$ we crossover to the localized regime. The corresponding
distributions (for $W=1.4$ and $W=2.4$) are shown in the inset of Fig. \ref{tau-dist2}. The
$\tau_{mp}$ in this case is proportional to the typical length traversed
by the particles that is the localization length $\xi$ \cite{amj}. Since
$\xi$ decreases with increasing $W$ so does $\tau_{mp}$.  The distribution
shows a long-time-tail due to the presence of resonances. For these
resonant realizations the particle explores the whole of the sample before
being totally reflected. Also, for this case one can notice that the 
distribution becomes narrow and weight at the tails becomes smaller with
increasing disorder in contrast with the behaviour observed for the tails
in the multiple scattering domain.

\begin{figure}[t] 
\protect\centerline{\epsfxsize=3in \epsfbox{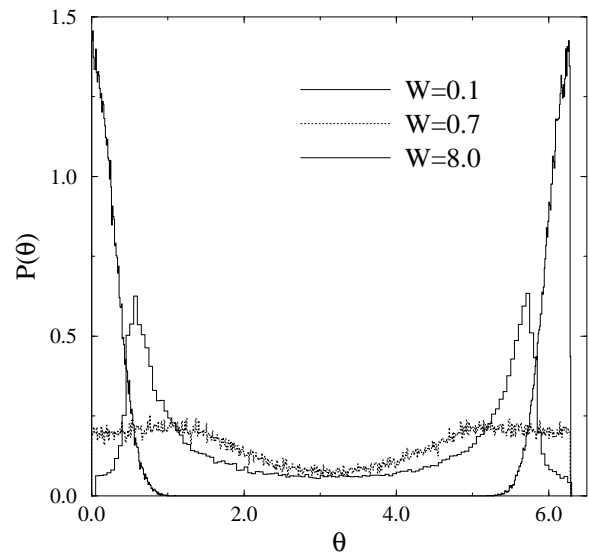}}
\caption{ Distribution of phase of reflected wave from the semi-infinite 
disordered sample of length $L=50$ for different values of disorder
strength $W$ in the different regimes namely, ballistic($W=0.1$), multiple
scattering ($W=0.7$) and strong localization ($W=8.0$).}
\label{PhaseDist} \end{figure}

In the ballistic regime the transport is dominated by few scattering
events and since the disorder is weak the phase changes associated with
these are small. Consequently, the distribution of phase of the reflected
wave shows (Fig. \ref{PhaseDist}) a doubly peaked structure in this
regime. The peaks occur at $0$ and $2\pi$. In the multiple scattering
regime the multiple scattering events provide ample opportunity for the
phase to get randomized. This is reflected in the fact that the
distribution shown in Fig. \ref{PhaseDist} for $W=0.7$ (multiple
scattering) is fairly uniform over the range $0$ to $2 \pi$. For $W=8.0$
we are in the strong localization regime and once again the double peak
structure emerges in the phase distribution. In this case, though the
double peak structure arises because $\xi \ll L$ i.e., the scattering is
so strong that most of the particles are reflected back from near the
surface thus getting scattered only from a few sites. In the limit of $W
\rightarrow \infty$ the peaks shift towards $\pi/2$ and $3\pi/2$.

\begin{figure}[t] 
\protect\centerline{\epsfxsize=3in \epsfbox{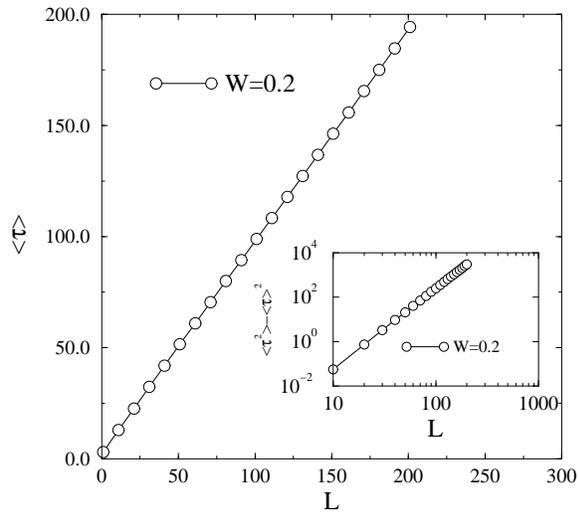}}
\caption{Plot of Average delay time $\avg{\tau}$ versus length of the
sample $L$ for disorder strength $W=0.2$. In the inset is the plot of
second cumulant of $\tau$ ($\avg{\tau^2}-\avg{\tau}^2$) versus lenth of
the sample $L$ for $W=0.2$. }
\label{Tau_vs_L}
\end{figure}

In Fig. \ref{Tau_vs_L} we plot the average value of time delay
$\avg{\tau}$ as a function of length $L$ of the sample. As predicted by
Texier et. al. \cite{texier} we see that the dependence is linear on $L$.
In the inset of the figure we plot, on log-log scale, the second cumulant
of $\tau$ $\avg{\tau^2} - \avg{\tau}^2$ as a function of the length $L$ of
the sample. We find a $L^{3.56}$ dependence of the second cumulant of
$\tau$ on the length of the sample in contrast to the $L^3$ behaviour
predicted by Texier et. al. \cite{texier}. The difference, probably, can
be attributed to the approximations involved in the analytical
calculations or the constraint of averaging over finite number of
realizations in our numerical calculations.

\section{Conclusion}
\label{conclude}

In summary, our study based on the statistics of Wigner delay time
distribution, shows that three qualitatively different regimes of
transport or scattering exist in disordered finite one-dimensional
systems. The systems exhibit a successive crossover from ballistic regime
to multiple scattering regime to localization regime as the strength of
disorder is varied.

%


\end{document}